\documentclass[a4paper]{jpconf}
\usepackage{graphicx}
\usepackage{wrapfig}
\usepackage{amsmath}
\begin{document}
\title{PICOLON dark matter search project}

\author{K.~Fushimi$^{1}$, 
D.~Chernyak$^{2}$, H.~Ejiri$^{3}$, K.~Hata$^{4}$, R.~Hazama$^{5}$, T.~Iida$^{6}$, H.~Ikeda$^{4}$, K.~Imagawa$^{7}$,
K.~Inoue$^{4,8}$, H.~Ishiura$^{9}$, H.~Ito$^{10}$, T.~Kishimoto$^{11}$, M.~Koga$^{4,8}$, K.~Kotera$^{12}$, 
A.~Kozlov$^{13}$, K.~Nakamura$^{14}$, R.~Orito$^{1}$, T.~Shima$^{3}$, Y.~Takemoto$^{8,15}$, S.~Umehara$^{3}$,
Y.~Urano$^{12}$, Y.~Yamamoto$^{1}$, K.~Yasuda$^{7}$, and S.~Yoshida$^{11}$
}
\address{$^{1}$ Graduate School of Technology, Industrial and Social Sciences, Tokushima University, 2-1 Minami Josanajima-cho, Tokushima city, Tokushima , 770-8506, Japan\\
$^{2}$ Department of Physics and Astronomy, University of Alabama, Tuscaloosa, Alabama 35487, USA \\
$^{3}$ Research Center for Nuclear Physics, Osaka University, 10-1 Mihogaoka Ibaraki city, Osaka, 567-0042, Japan\\
$^{4}$ Research Center for Neutrino Science, Tohoku University, 6-3 Aramaki Aza Aoba, Aobaku, Sendai city, Miyagi, 980-8578, Japan\\
$^{5}$ Department of Environmental Science and Technology, Osaka Sangyo University, 3-1-1 Nakagaito, Daito city, Osaka, 574-8530, Japan\\
$^{6}$ Faculty of Pure and Applied Sciences, University of Tsukuba, 1-1-1 Tennoudai, Tsukuba city, Ibaraki, 305-8571, Japan\\
$^{7}$ I.~S.~C. Lab.~, 5-15-24 Torikai Honmachi, Settsu city, Osaka, 566-0052, Japan\\
$^{8}$ Kavli Institute for the Physics and Mathematics of the Universe (WPI), 5-1-5 Kashiwanoha, Kashiwa city, Chiba, 277-8583, Japan\\
$^{9}$ Department of Physics, Graduate School of Science, Kobe University, 1-1 Rokkodai-cho, Nada-ku, Kobe city, Hyogo, 657-8501, Japan\\
$^{10}$ Department of Physics, Faculty of Science and Technology, Tokyo University of Science, Noda, Chiba 278-8510, Japan\\
$^{11}$ Department of Physics, Osaka University, 1-1 Machikaneyama-cho, Toyonaka city,  Osaka 560-0043, Japan\\
$^{12}$ Graduate School of Advanced Technology and Science, Tokushima University, 2-1 Minami Josanajima-cho, Tokushima city, Tokushima , 770-8506, Japan\\ 
$^{13}$ National Research Nuclear University ``MEPhI'' (Moscow Engineering Physics Institute), Moscow, 115409, Russia\\
$^{14}$ Butsuryo College of Osaka, 3-33 Ohtori Kitamachi, Nishi ward, Sakai city, Osaka, 593-8328, Japan\\
$^{15}$ Institute for Cosmic Ray Research, The University of Tokyo, 5-1-5 Kashiwanoha, Kashiwa city, Chiba, 277-8583, Japan\\
}
\ead{kfushimi@tokushima-u.ac.jp}

\begin{abstract}
PICOLON (Pure Inorganic Crystal Observatory for LOw-energy Neutr(al)ino) aims to search for cosmic dark
matter by high purity NaI(Tl) scintillator.
We developed extremely pure NaI(Tl) crystal by hybrid purification method. The recent result of 210Pb in our
NaI(Tl) is less than 5.7 $\mu$Bq/kg.
We will report the test experiment in the low-background measurement at Kamioka Underground Laboratory.
The sensitivity for annual modulating signals and finding dark matter particles will be discussed.
\end{abstract}

\section{Introduction}
The search for WIMPs (Weakly Interacting Massive Particles) is crucial in determining candidates for cosmic dark matter.
In the 21st century, many groups searched for WIMPs using various target nuclei, resulting in null results.
Only the DAMA/LIBRA group with a 250 kg NaI(Tl) detector reported an annually modulating signal between
2 keV$_{\mathrm{ee}}$ and 6 keV$_{\mathrm{ee}}$.
Where keV$_{\mathrm{ee}}$ is the electron equivalent energy of the recoil nucleus calibrated by the kinetic energy of the electron.

In many experiments with other target nuclei, the region of WIMPs mass and scattering cross-sections corresponding to
annual modulation has been denied.
However, only the NaI (Tl) detector of DAMA/LIBRA has observed a significant seasonal variation of $12.9 \sigma$ for a period of 13 cycles or more.
It is necessary to verify whether this is due to the equipment of the DAMA/LIBRA group or a peculiar phenomenon observed only in NaI (Tl).

The COSINE-100 and the ANAIS-112 are leading the way in verification experiments using NaI (Tl).
Although their background is $2\sim4$ times larger than the DAMA/LIBRA group,
they have investigated the origin of the background precisely by long-term measurement and succeeded
in explaining most of the background by simulation \cite{Park_Poster}.

The most reliable verification is to perform experiments using a highly sensitive NaI (Tl) detector
with a low background equal to or less than the DAMA/LIBRA group.
Preventing this is the radioactive impurities of $^{210}$Pb and $^{40}$K contained in NaI (Tl) crystals.
So far, all NaI (Tl) crystals have been contaminated with high-concentration impurities of several hundred $\mu$Bq/kg or more.
Consequently, the background could not be reduced sufficiently.
The Table \ref {tb:genjou} lists the concentrations of radioactive impurities in NaI (Tl) crystals currently reported by each group.
\begin{table}[ht]
\centering
\caption{Present status of the radioactive contamination and background level in NaI(Tl) groups.
$^{210}$Pb, $^{226}$Ra, and $^{228}$Th are in unit of $\mu$Bq/kg assuming the radioactive equilibrium.
$^{\mathrm{nat}}$K is in unit of ppb. The BG level is in unit of day$^{-1}$keV$^{-1}$kg$^{-1}$.}
\label{tb:genjou}
\begin{tabular}{l|rrrrr|c} \hline
Group & $^{210}$Pb & $^{226}$Ra & $^{228}$Th & $^{\mathrm{nat}}$K & BG & Reference \\ \hline
COSINE-100 & $25\pm5$ & $7\pm2$ & $<20$ & $<42$ & $2\sim3$ & \cite{Park_Poster}\\
ANAIS-112 & $700\sim3150$ & $2.7\sim10$ & $0.4\sim4$ & $18\sim44$ & $3\sim4$ & \cite{Cebrian_Poster} \\
SABRE & $410\pm20$ & $5.9\pm0.6$ & $1.6\pm0.3$ & 20 & $4.6\pm0.2$ & \cite{Mariani_Poster} \\ \hline
\end{tabular}
\end{table}

The PICOLON (Pure Inorganic Crystal Observatory for LOw-energy Neutr (al) ino) group investigated the method to remove radioactive impurities. We established the optimum combination and succeeded in reducing the background.

\section{Purification}
Establishing a purification method is essential for making a highly sensitive detector.
We searched for the optimum combination of the recrystallization method for removing the radioactive impurities.

As reported by COSINE-100, the recrystallization method is very effective in removing water-soluble impurities, for example, potassium \cite{COSINE_Purify}.
However, since many compounds of Pb, Ra, and other heavy isotopes are not water-soluble,
it is necessary to find an effective method other than the recrystallization method.
We have tried several ion exchange resins and found the optimal combination.
As a result, we succeeded in significantly removing radioactive impurities that emit alpha rays such as
$^{210}$Pb.
As shown in the Table \ref{tb:history}, the purity of NaI (Tl) by our project so far has been improved by
optimizing the purification method.
The detailed information of the purification is described in the previous paper \cite{Fushimi_PTEP}.

\section{Low background measurement}
\subsection{Data acquisition method and analysis}
Low background measurements were performed at the Kamioka Underground Laboratory in Hida City, Gifu Prefecture, JAPAN.
The laboratory is located in the KamLAND area of
2700~m.~w.~e.~ (meter water equivalent) underground.
In July 2021, we installed two NaI (Tl) modules, Ingot \#85 and Ingot \#94, in a passive shield that
consists of 20 cm thick lead and a 5 cm thick OFHC (Oxygen-Free High conductivity) copper.
We installed them in different shields to measure the individual backgrounds.

A cylindrical NaI (Tl) crystal 76.2~mm in diameter and 76.2 ~m in length is covered with an ESR$^{\mathrm{TM}}$ reflective sheet
whose thickness is 65~$\mu$m thick.
The crystal is sealed in a 3~mm thick acrylic housing with a synthetic quartz optical window on one end.
The scintillation light is converted into a current signal by R11065-20mod photomultiplier tube (PMT) made by Hamamatsu Photonics.

\begin{figure}[ht]
\centering
\includegraphics[width=0.52\linewidth]{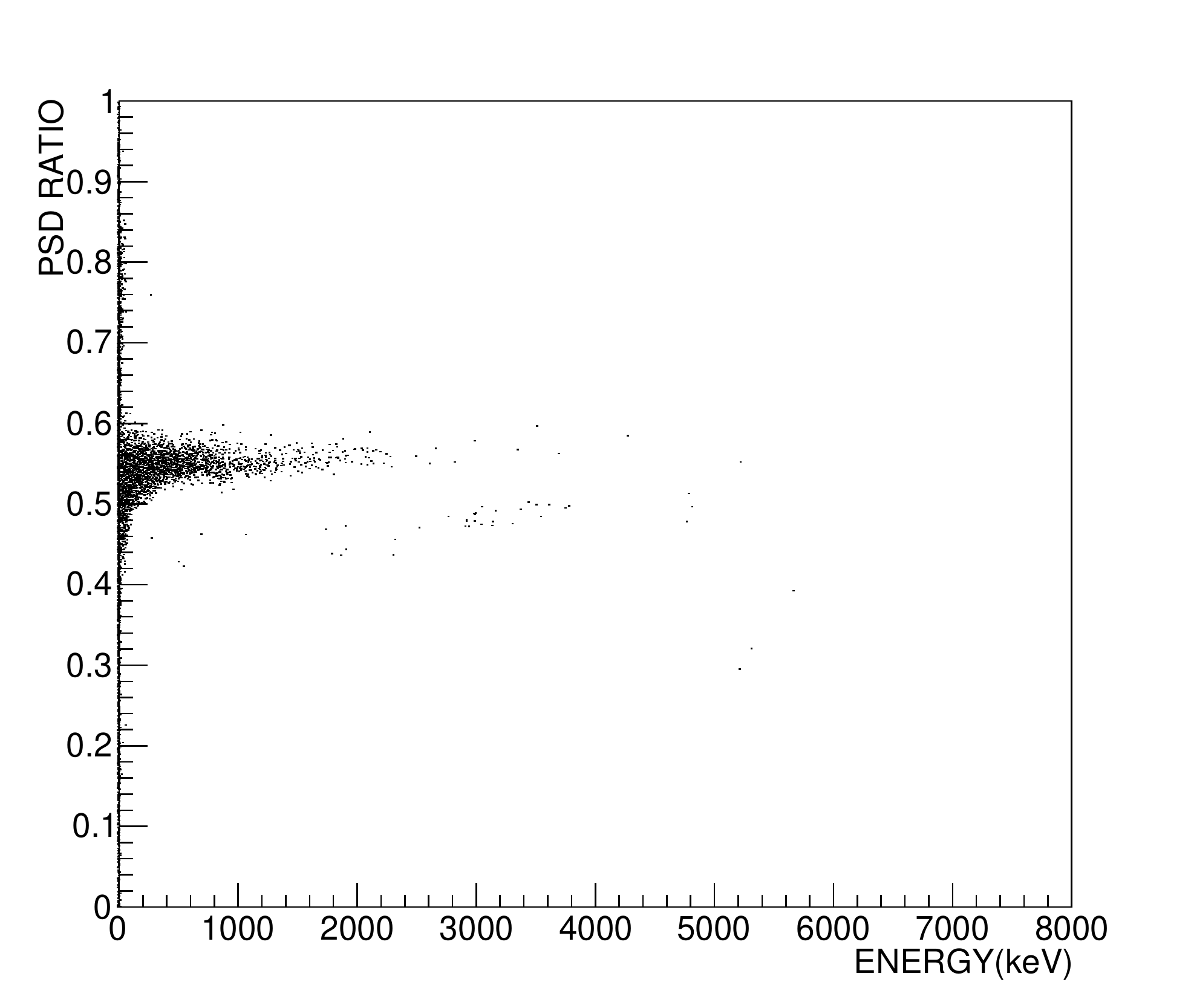}
\includegraphics[width=0.45\linewidth]{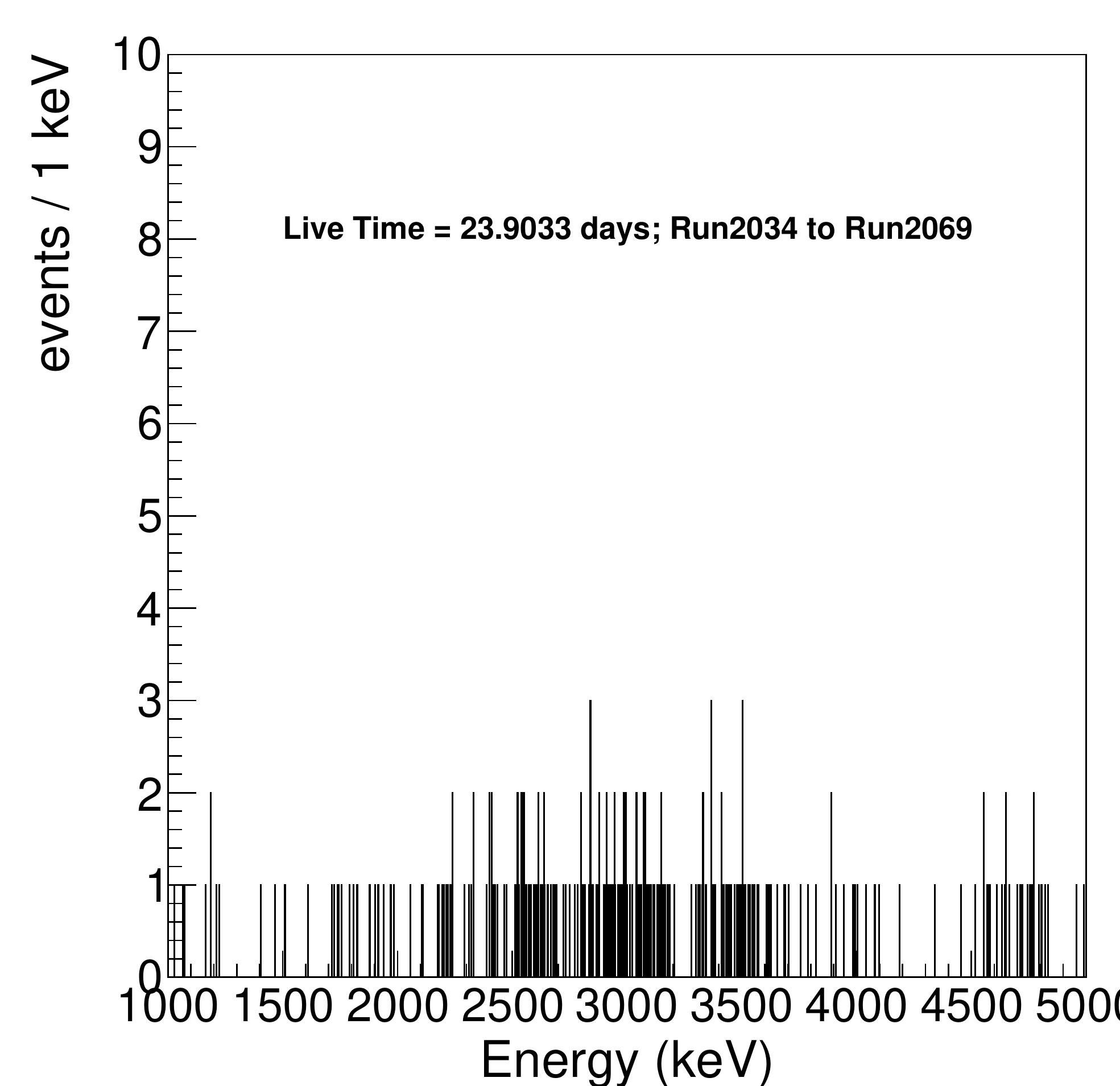}
\caption{Left: The PSD scatter plot of Ingot \#94. Right: The energy spectrum of alpha rays in
Ingot \#94.}
\label{fg:I94}
\end{figure}

The waveform of the signal was acquired by the high-speed analog-to-digital converter MoGURA.
Alpha ray counting and the analysis of low energy were performed by offline analysis.
\subsection{Alpha ray analysis}
The scintillation decay time of NaI(Tl) is 190~nsec for alpha rays and 230~nsec for electrons.
Let $R$ be the ratio of the value obtained by integrating all the current at the PMT output to the value obtained by integrating only after 200~nsec after the start of the signal.
Since the value of $R$ is smaller for the alpha ray event, it is possible to extract the alpha ray events from the singles events\cite{Ichihara_PSD}.

The left panel of the Figure \ref{fg:I94} shows a scatter plot;  $R$ on the vertical axis, and the energy on the horizontal axis.
The beta/gamma ray region and the alpha ray region are separated.
In the high-energy region above 5000~keV, the PSD plot is bent downward. 
This bent is due to the distortion caused by the large output signal of the PMT. 
We discussed with Hamamatsu Photonics and adjusted the parameters of the PMT divider circuit 
so that the signal distortion would not affect the peak structure of the alpha-ray.

\begin{table}[ht]
\centering
\caption{The purification history of our NaI(Tl) ingots.}
\label{tb:history}
\begin{tabular} {l|rrr} \hline
Ingot & \#73 & \#85 & \#94 \\ \hline
$^{\mathrm{nat}}$K (ppb) & $<30$ & $<20$ & $<20$ \\
$^{232}$Th (ppt) & $1.8\pm0.2$ & $0.3\pm0.5$ & $<6$ \\
$^{238}$U (ppt) & $9.4\pm0.8$ & $1.0\pm0.4$ & $<2$ \\
$^{210}$Pb ($\mu$Bq/kg) & 1300 & $<5.7$ & $<6$ \\
Method & RC$\times 3$ & RC$\times 2$+Resin & RC$\times 2$+Resin \\ \hline
\end{tabular}
\end{table}

The energy spectrum of alpha rays is shown in the right panel of Figure \ref{fg:I94}.
Despite the insufficient statistical accuracy due to the short live-time, we have succeeded in reproducing the measurement results in 2020.
The Table \ref{tb:history} shows the concentrations of radioactive impurities.
The purification methods are also listed.
Ingot94 also achieved sufficiently high purity, confirming that a purification method was established.

\subsection{Low energy spectrum}
The background due to noise events is the main component in the energy region lower than
a few tens of keV or less.
Most of the noise events consist of PMT dark currents and Cherenkov radiation.
Since both time constants are sufficiently shorter than those of NaI (Tl),
PSD can distinguish them from NaI (Tl) signals.
The background event was identified from the NaI (Tl) signal by whether the next signal
came within 200 nsec after the first signal appeared.
In the case of dark current or Cherenkov radiation, no following signal comes except for
the first occurrence, but in the case of NaI (Tl) scintillation,
many signals are generated during 200 $\mu$sec.
\begin{wrapfigure}{r}{0.5\linewidth}
\centering
\includegraphics[width=\linewidth]{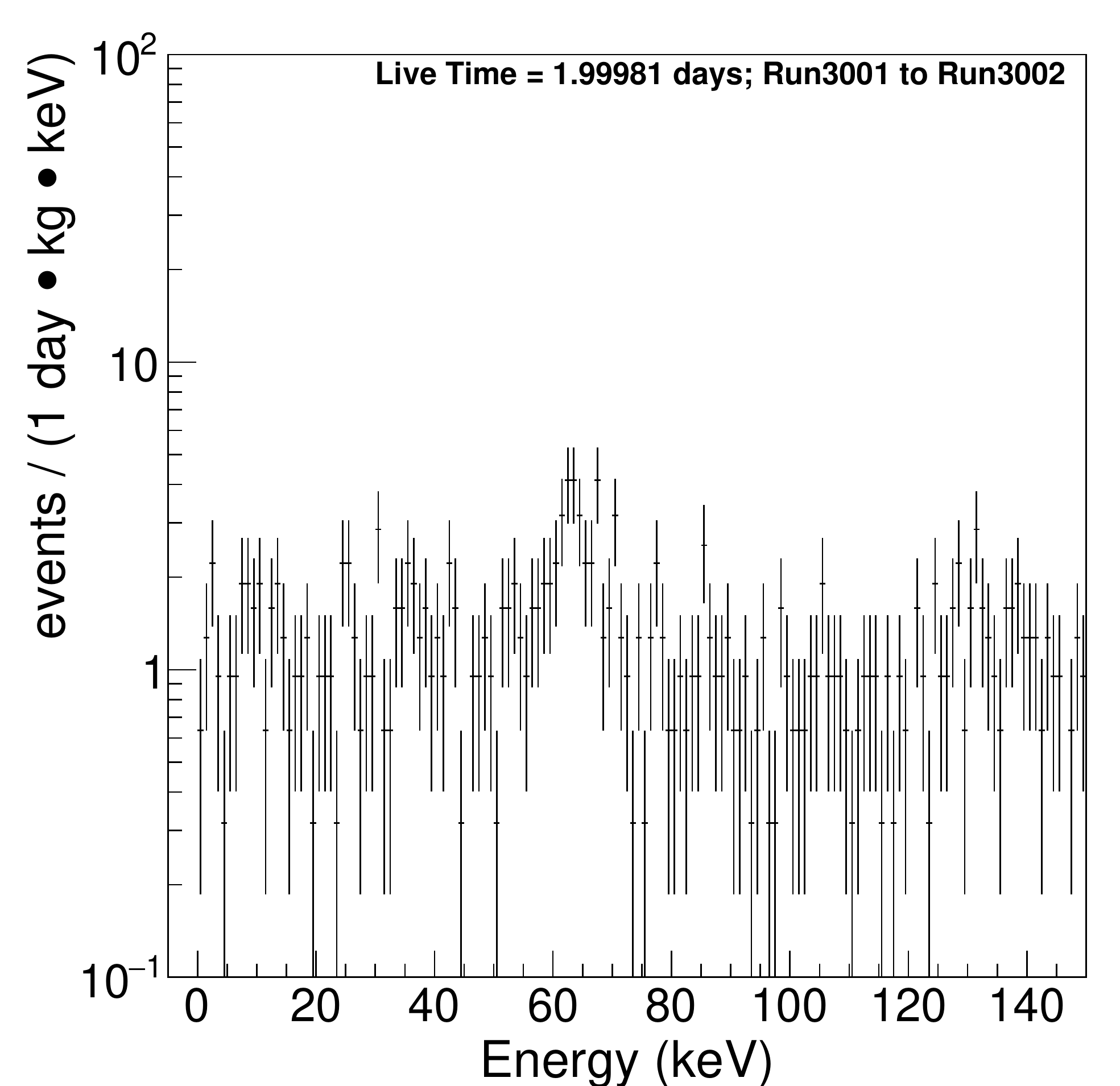}
\caption{The low energy spectrum taken from Ingot\#94.}
\label{fg:lowe}
\end{wrapfigure}

Figure \ref{fg:lowe} shows the energy spectrum of the low energy region with the background removed.
Significant peaks have been observed near 35 keV$_{\mathrm{ee}}$ and 65 keV$_{\mathrm{ee}}$.
These are the characteristic X-rays and gamma rays emitted
by the cosmogenic nuclide $^{125}$I by electron capture.

\section{Prospects}
We have established a purification method for NaI(Tl) and succeeded in producing 
ultra-pure crystals necessary to search cosmic dark matter.
We have succeeded in achieving high purity of 10 $\mu$Bq/kg for $^{226}$Ra and $^{232}$Th,
 and less than 600 $\mu$Bq/kg for $^{40}$K.
Notably, the concentration of $^{210}$Pb was ultrapure, less than 5.6 $\mu$Bq/kg.

The background counting rate by only one NaI(Tl) detector module was 1.27~dru in the range of 1$\sim$10~keV$_{\mathrm{ee}}$. This value is the lowest background count rate for NaI(Tl) detectors without anti-coincidence.

In the fiscal year of 2021, we will construct several large NaI(Tl) crystals of 12.7 cm in diameter and 12.7 cm in length and perform an anti-coincidence measurement.
The background level is expected to be reduced by a factor of a few by anti-coincidence measurement between the modules.
It is expected that the NaI(Tl) detector, which has a lower background than that of the DAMA/LIBRA group, will provide a reliable verification.

\begin{wrapfigure}{r}{0.5\linewidth}
\centering
\includegraphics[width=\linewidth]{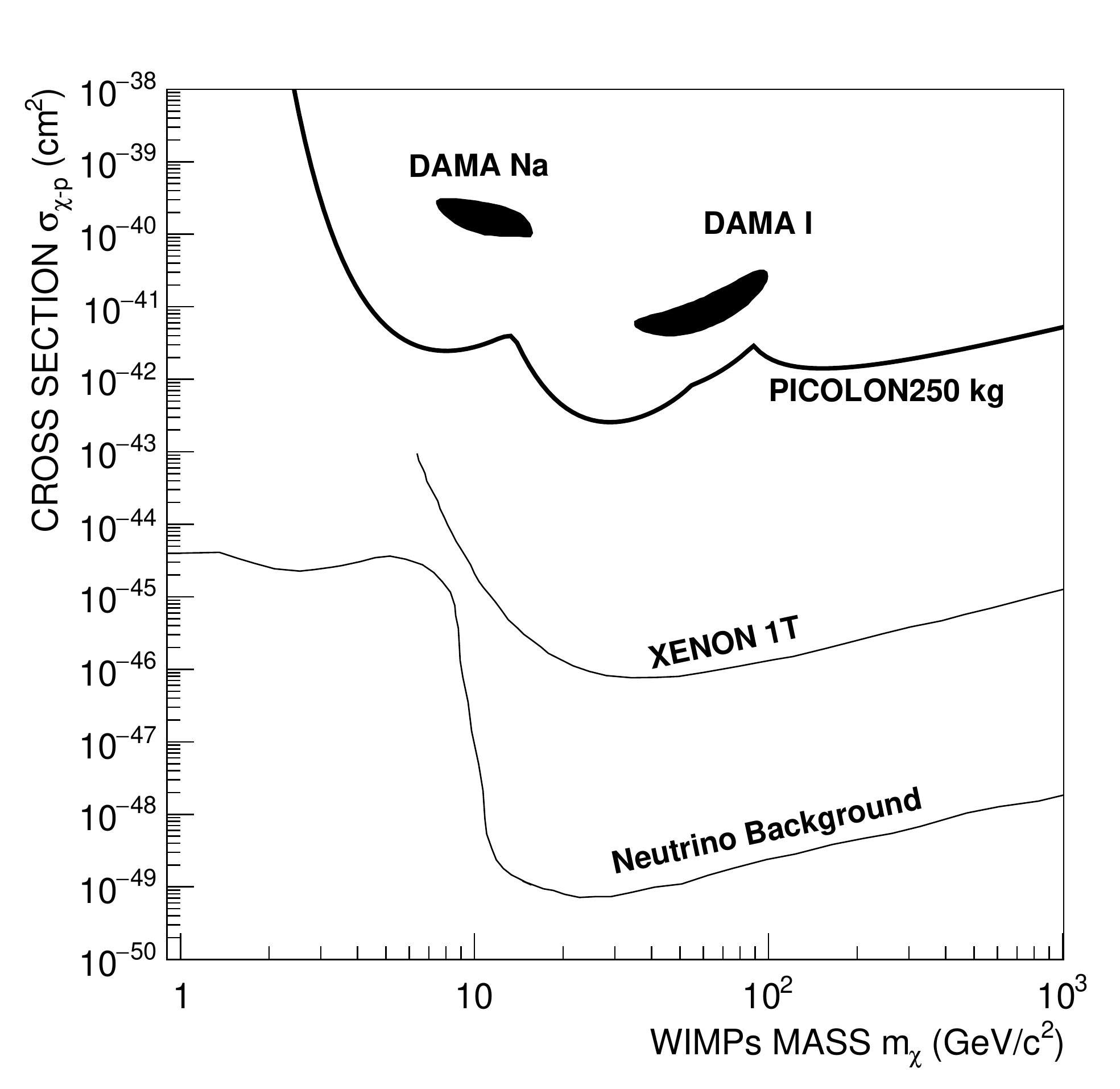}
\caption{The expected sensitivity to spin-independent WIMPs.}
\label{fg:furoshiki}
\end{wrapfigure} 
The PICOLON project is planning to construct 250~kg of highly radiopure NaI(Tl) detector.
It will consist of 42 modules of NaI(T) detectors.
The dimension of the highly radiopure NaI(Tl) is 12.7~cm in diameter and 
12.7~cm in length.
The expected background rate is 1~/day/keV/kg at 1~keV$_{\mathrm{ee}}$, which is the modest value.
The expected sensitivity to the project is shown in Figure~\ref{fg:furoshiki}.

As shown in Figure ~\ref{fg:furoshiki},  our experiment can verify the result of 
the DAMA/LIBRA experiment after one year of continuous data taking.

\section{Acknowlegement}
We acknowledge the support of the Kamioka Mining and Smelting Company. 
This work was supported by JSPS KAKENHI Grant No. 26104008, 19H00688, 20H05246, 
and Discretionary expense of the president of Tokushima University. 
This work was also supported by the World Premier International Research Center
Initiative (WPI Initiative).

\end{document}